\documentclass[aip,jap,graphicx,reprint]{revtex4-1}
\usepackage{amsmath}
\usepackage{amssymb}
\usepackage{graphicx}
\usepackage{hyperref}
\usepackage{times}

\begin{document}

\title{Simulation and modeling of the electronic structure of GaAs damage clusters}

\author{Jonathan E. Moussa}

\author{Stephen M. Foiles}
\email{foiles@sandia.gov}

\author{Peter A. Schultz}

\affiliation{Sandia National Laboratories, Albuquerque, NM 87185, USA}

\date{\today}

\begin{abstract} \centering \begin{minipage}{0.79\textwidth}
In an effort to build a stronger microscopic foundation for radiation damage models in gallium arsenide (GaAs),
 the electronic properties of radiation-induced damage clusters are studied with atomistic simulations.
Molecular dynamics simulations are used to access the time and length scales required for direct simulation of a collision cascade,
 and density functional theory simulations are used to calculate the electronic properties of isolated damaged clusters
 that are extracted from these cascades.
To study the physical properties of clusters,
 we analyze the statistics of a randomly-generated ensemble of damage clusters
 because no single cluster adequately represents this class of defects.
The electronic properties of damage clusters are accurately described by a classical model of the electrical charging
 of a semiconducting sphere embedded in an uniform dielectric.
The effective band gap of the cluster depends on the degree of internal structural damage,
 and the gap closes to form a metal in the high-damage limit.
We estimate the Fermi level of this metallic state, which corresponds to high-energy amorphous GaAs,
 to be $0.46 \pm 0.07$ eV above the valence band edge of crystalline GaAs.
\end{minipage} \end{abstract}

%61.43.Bn	Structural modeling: serial-addition models, computer simulation (disordered solids)
%71.15.Mb	Density functional theory, local density approximation, gradient and other corrections
%71.55.Eq	III-V semiconductors (impurity and defect levels)
\pacs{61.43.Bn,71.15.Mb,71.55.Eq}

\maketitle

\section{Introduction}

%P1.1 - Basic motivation, ``multiscale''
Modeling the effect of radiation damage on an electronic circuit is a multiscale effort.
Circuit simulations \cite{xyce} rely on compact damage models \cite{compact},
 which are constructed from device simulations of single circuit elements \cite{charon}.
In turn, electronic device simulation relies on defect models \cite{defect_model}
 that include the drift, diffusion, creation, and annihilation of defects
 with appreciable concentrations and non-negligible electronic activity.
Identification and characterization of such defects requires atomistic simulations.
These simulations are further divided into multiple scales and categorized by the choice of microscopic degrees of freedom.
Binary collision approximation (BCA) simulations \cite{BCA} only track atoms that deviate from the ideal crystal structure
 and are only limited by the total amount of crystal damage simulated rather than simulation volume.
Classical molecular dynamics (MD) simulations \cite{LAMMPS} track all atoms in a simulation volume,
 which restricts the simulation size while more accurately accounting for interatomic interactions.
Density functional theory (DFT) simulations \cite{VASP} include a quantum description of electrons
 and enable the simulation of charge fluctuations, but with an asymptotic cost that scales cubicly with simulation volume.
Fidelity of atomistic simulations is particularly important because microscopic errors can propagate
 through the simulation scales into macroscopic errors.

%P1.2 - Target: GaAs. Traditional approach based on point defects.
We consider the case of gallium arsenide (GaAs), where modeling is necessary
 because of a relative dearth of experimental characterization of defects and radiation damage
 compared to silicon.
The central assumption of radiation damage modeling at the device scale is that damage can be reduced to
 a time-dependent spatial distribution of point defects and their binary chemistry
 with electrons, holes, and other defects.
In accordance, recent DFT studies have characterized the electronic properties
 of all likely point defects in GaAs \cite{Peter1,Peter2}.
The point-defect ansatz is consistent with the BCA of high energy collision cascades that consist primarily of well-separated Frenkel pairs,
 but it breaks down at the end of a collision cascade when a particle loses sufficient energy that the mean distance
 between collisions approaches the interatomic separation (the ``end of range'').
MD simulations of this regime are consistent with the BCA in accounting for populations of point defects,
 but they also contain damage clusters with a comparable number of atoms
 that cannot be decomposed into a collection of isolated point defects \cite{MD_damage}.

%P1.3 - Experimental connections
The end of range in an MD simulation is the spatial region surrounding the terminus of a single collision cascade event.
Standard experiments do not resolve a single cascade event and the end of range instead refers to the average depth at which cascades terminate.
The end-of-range layer in ion implantation experiments contains the highest density of material damage
 and nanoscale clustering of defects is observed regularly \cite{GaAs_EOR}.
The damage to GaAs caused by ion implantation and neutron irradiation produces broad anomalous features (the `U' and `L' bands) in
 deep-level transient spectroscopy (DLTS) \cite{GaAs_DLTS}, which is an experimental probe used to characterize the energy levels of point defects.
Such features are absent in pristine and electron-irradiated GaAs.
Experimental characterization of end-of-range damage remains an open problem and will benefit from
 microscopic insights derived from theory and simulation.

%P1.4 - Modeling of damage clusters
The structural and electronic properties of GaAs damage clusters have not been studied
 to assess their importance in radiation damage models relative to point defects.
For point defects in GaAs, there are a manageable number of plausible structures
 that can be enumerated and simulated exhaustively.
In contrast, damage clusters can contain a large number of atoms and span a very large configuration space.
No single structure would be adequately representative or predictive of the properties of damage clusters.
Instead, a study of their properties requires a statistical analysis of many cluster configurations.
 
%P1.5 - Summary
In this paper, we use a combination of MD and DFT simulations to model the electronic properties
 of damage clusters in GaAs and strengthen the microscopic foundation upon which future radiation damage models will be built.
First, we collect a sample set of damage clusters generated from MD simulations of collision cascades.
Because of the discrepancy in computationally tractable simulation volumes between MD and DFT,
 the sampling is biased towards smaller clusters that are accessible to DFT calculations.
We supplement this set with a larger sample set of artificially generated idealized damage clusters
 that are uniformly distributed over a range of cluster sizes.
This artificial set is large enough to generate reliable statistics, which enables us to carefully assess finite-size effects
 and extrapolate models into a regime of cluster size that is not directly accessible with DFT.
To complement the extrapolation, large damage clusters are studied indirectly by generating bulk amorphous GaAs structures
 that are assumed to be representative of the cluster interior.
The atomistic simulation methodology is reviewed in Section \ref{method_section}.
Several models of the electronic properties of defect clusters are proposed in Section \ref{model_section}.
The results of simulation and model fitting are reported in Section \ref{result_section},
 and their implications for device-level radiation modeling are discussed in Section \ref{discuss_section}.

\section{Computational methodology\label{method_section}}

%P2.1 - Basic MD methodology (LAMMPS): potential, box, initial conditions, time length, time step (Foiles)
The MD simulations are performed using the \textsc{Lammps} \cite{LAMMPS} code.
They employ a Pettifor bond-order potential for GaAs developed by Murdick and co-workers \cite{GaAs_pot}
 that has been modified at small interatomic separations to match the Ziegler-Biersack-Littmark (ZBL) universal repulsive potential \cite{ZBL}.
The basic methodology for simulating displacement cascades is well-established and has been reviewed by Averback and de la Rubia \cite{Averback}.
In order to simulate a neutron-induced recoil, an atom is selected at random and given an impulse in a random direction with the desired energy.
The system is evolved for a period of 10 ps using a variable time step which ensures that no atom moves more than 1 pm in a given step.
The size of the simulation cell depends on the recoil energy to be simulated.
The simulations of 1 keV recoils contain 216,000 atoms, while the simulations of 50 keV recoils contain 13,824,000 atoms.
More details of these simulations are described elsewhere \cite{Foiles_SAND, Foiles_prep}.

%P2.2 - Amorphous zone identification (Foiles)
Regions of GaAs that cannot be decomposed into point defects embedded in crystalline GaAs are categorized as amorphous regions.
They are identified in the MD simulations using a generalization of an approach
 used earlier by Foiles \cite{MD_damage} to analyze displacement cascades in silicon.
The criteria used to identify the amorphous regions is the presence of 5- and 7-atom rings.
A ring in this context is a path along nearest-neighbor atoms that returns to its original site.
In the case of an ideal diamond structure, the smallest rings contain 6 or 8 atoms along the path.
To define an amorphous region, all rings of 7 atoms or less are identified.
The system is divided into cells that correspond to the cubic cell of the diamond structure,
 and the number of 5- and 7-atom rings centered in each cubic cell is determined.
Cells that contain more than one 5-atom ring or more than two 7-atom rings are classified as amorphous regions.

%P2.3 - Damage cluster extraction (Foiles)
The DFT simulation volume used in this study is a $4 \times 4 \times 4$ supercell of the conventional cubic unit cell
 of GaAs (a 11.2 nm$^3$ cube) containing 256 Ga and 256 As atoms.
Damage clusters are required to be contained within this volume with approximately 1 nm of crystalline GaAs
 as a buffer between the damage cluster and its periodic image.
Amorphous regions that satisfy this criterion are visually identified and extracted from the MD simulations.
Ten damage clusters are generated using this procedure and then relaxed with MD to their metastable ground states.
In three cases, the extracted subsystem is further annealed with MD for 100 ns at 300 K.

%P2.4 - Amorphous bulk generation (Foiles)
Two bulk amorphous GaAs structures are also creating using MD simulation,
 starting from a $4 \times 4 \times 4$ cubic supercell of the ideal crystal.
The crystal is heated to 2500 K and isothermally evolved for 1 ns, then the temperature is reduced
 to 300 K over 15 ns and isothermally evolved for 10 ns.
This process produces a large number of Ga-Ga and As-As nearest neighbors,
 which are energetically unfavorable relative to Ga-As nearest neighbors.
A second, lower-energy amorphous structure is created from the first structure by applying a Monte Carlo procedure
 to pairwise exchange atoms in order to maximize the number of Ga-As nearest neighbors.
This structure is heated to 800 K, cooled to 300 K over 5 ns and isothermally evolved for 10 ns.

%P2.5 - Generation of artificial clusters
Artificial damage clusters are generated using a simple procedure to simulate high-energy damage.
We again begin the process with $4 \times 4 \times 4$ cubic supercells of the ideal GaAs crystal.
Within a sphere of a given radius that is arbitrarily centered on an As atom,
 all atoms on lattice sites and all vacancies on interstitial sites are randomly assigned to new sites, lattice or interstitial, within the sphere.
5 different radii are chosen -- 8, 10, 12, 14, and 16 bohr (1 bohr $\approx$ 0.0529 nm) --
 and 25 clusters are generated for each of these radii, for a total of 125 artificial clusters.

%P2.6 - Basic DFT methodology (QUEST)
DFT calculations of the damage clusters are performed with the \textsc{SeqQuest} code \cite{quest}.
The details of these calculations are guided by previous studies of point defects in GaAs \cite{Peter1}.
The local density approximation (LDA) is used for electron exchange and correlation \cite{PZ81}.
Calculations using PBE \cite{PBE} for simple intrinsic defects do not exhibit any physically meaningful differences \cite{Peter1}.
Norm-conserving pseudopotentials without semicore $d$-orbitals but with nonlinear core corrections are used for both Ga and As atoms \cite{pp_GaAs},
 and a double-zeta-plus-polarization quality atomic orbital basis set is used to represent crystal orbitals.
The Brillouin zone is sampled only at the $\Gamma$ point for the $4 \times 4 \times 4$ supercells,
 sufficient for the accuracy needed in the current study.
Gaussian smearing of 0.03 eV is applied to orbital occupations to aid in convergence of the self-consistent field cycle
 in the presence of multiple energy levels near the Fermi level.
The lattice constant is fixed at the LDA value for GaAs, 0.560 nm.
All damage clusters are relaxed to the DFT local energy minimum using an accelerated steepest descent 
 update of atomic positions to a tolerance of 0.1 eV/nm.
Vertical charge transition energies are calculated by constraining the structure
 of the charged damage clusters to the neutral geometries.
Adiabatic charge transition energies are calculated by relaxing the structure of each charge state
 to its local energy minimum starting from the relaxed geometry
 of the previous charge state with one fewer unit of charge. 

%P2.7 - Further details of the local moment charge method
DFT calculations of charged damage clusters use the local moment countercharge (LMCC) method \cite{countercharge}.
The charge density in the unit cell is partitioned into a Gaussian approximation of the defect charge plus a charge-neutral remainder.
The electrostatic potential within the unit cell is constructed from the periodic charge neutral remainder and the nonperiodic Gaussian defect charge.
The total energy is corrected with an embedding energy to align the charged defect to the perfect crystal potential
 and a bulk screening energy to account for polarization outside the unit cell.
A supercell procedure is used to account for bulk screening energy outside the volume of the unit cell,
 which was calibrated to converge to the bulk dilute limit in GaAs with a 1.6 bohr skin depth \cite{Peter1}.

%P2.8 - Amorphous bulk DFT calculations (VASP)
DFT calculations of the two bulk amorphous GaAs structures are performed with the \textsc{Vasp} code \cite{VASP}.
The details of these calculations are slightly different from the \textsc{SeqQuest} simulations.
The LDA is still used to relax the structure, but the lattice constant is kept at the experimental value of 0.565 nm.
Default projector augmented wave (PAW) pseudopotentials are used for both Ga and As,
 without $d$-orbitals in the valence (consistent with the \textsc{SeqQuest} calculations).
A different model of electron exchange and correlation, the Heyd-Scuseria-Ernzerhof (HSE06) model \cite{HSE06},
 is used for electronic density of states (DOS) calculations.
The resulting HSE06 band gap of crystalline GaAs is 1.26 eV, which compares reasonably well to the experimental value of 1.52 eV \cite{GaAs_gap}.
Because the HSE06 density functional produces accurate band gaps, the accompanying DOS
 can serve as a plausible representation of a photoemission spectrum.
Here, the Brillouin zone is sampled on a $2 \times 2 \times 2$ $\Gamma$-centered grid
 and a Gaussian smearing of 0.1 eV is applied to orbital occupations to reduce sampling artifacts in the DOS.
For comparison, the HSE06-based DOS of crystalline GaAs is computed in \textsc{Vasp} with high energy resolution
 using a single unit cell with a $32 \times 32 \times 32$ sampling of the Brillouin zone
 and tetrahedral Brillouin zone integration.
  
\section{Electronic models of damage clusters\label{model_section}}

%P3.1 - Introduction to damage modeling
We develop a damage cluster model in three stages of increasing physical complexity and number of free parameters.
The goal is to produce a total energy model of a damage cluster, $E(Q)$, as a function of its total charge, $Q$,
 relative to the uncharged state, $E(Q=0) \equiv 0$.
$Q$ is an integer in units of $|e|$, the magnitude of electron charge.
Given an electron chemical potential, $\mu$, the stable charge state of the cluster, $Q_S(\mu)$,
 is defined by minimizing the energy over the transfer of charge between the crystalline bulk and the cluster,
\begin{equation}
 E(Q_S(\mu)) + \mu Q_S(\mu) = \min_{Q \in \mathbb{Z}} \left[  E(Q) + \mu Q \right] .
\end{equation}
$Q_S(\mu)$ is usually positive when $\mu$ is near the valence band edge, $E_V$,
 and negative when $\mu$ is near the conduction band edge, $E_C$.

%P3.2 - Charge transfer energy levels
Charge transfer energy levels of the damage cluster, $\Delta^Q_{Q'}$, are the values of $\mu$
 at which two different charge states, $Q$ and $Q'$, have equal energy ($Q' > Q$ by convention).
$\Delta^Q_{Q'}$ has a simple analytic formula,
\begin{equation}
 \Delta^Q_{Q'} = \frac{E(Q) - E(Q')}{Q' - Q}.
\end{equation}
The most physically important energy levels are the transitions between stable charge states,
 where $Q = Q_S(\Delta^Q_{Q'} + \delta)$ and $Q' = Q_S(\Delta^Q_{Q'} - \delta)$
 for an infinitesimal energy $\delta$.
Typically, the transition between stable charge states is a single-electron process with $Q' = Q+1$.
If the transition is a multi-electron process, then all the single-electron $Q''$ to $Q''+1$ transitions
 involving metastable charge states between $Q$ and $Q'$ are also physically important
 because they occur at a higher rate than the multi-electron process.
The simplest case is $Q' = Q + 2$, which is characterized as negative-$U$ behavior \cite{negU}.
Here, the $\Delta^Q_{Q+2}$ transition between stable states is halfway between the $\Delta^Q_{Q+1}$
 and $\Delta^{Q+1}_{Q+2}$ energy levels.
We only consider damage cluster models with single-electron transitions between stable charge states,
 which is valid when structural relaxation effects are small.

%P3.3 - Minimal model based on arguments of minimal structure
The first stage of electronic model construction is a basic argument for minimal structure.
The energy levels of point defects can occur anywhere within the band gap between
 the valence band edge and conduction band edge of GaAs, which is an energy window of 1.52 eV \cite{GaAs_gap}.
While computed point defect levels in GaAs are observed to have some preference for the lower half of the band gap \cite{Peter1},
 levels were calculated to span the band gap (despite a nominal band gap problem in the LDA and PBE functionals),
 and we assume no bias in the larger configuration space of damage clusters.
If the distribution of levels within the gap is uniform, then it can be characterized on average by
 a capacitive total energy model that is a quadratic polynomial in charge,
\begin{align}\label{model1}
 E(Q) &= -\mu_0 Q + \tfrac{1}{2 C} Q^2 , \\
 \Delta^Q_{Q+1} &= \mu_0 - \tfrac{1}{C} (Q + \tfrac{1}{2}) \notag ,
\end{align}
 with two free parameters, $\mu_0$ and $C$, that set an origin and energy density for the defect levels.
As yet, the model does not have any microscopic physical content.

%P3.4 - Spherical metal inclusion model, 2 parameters
The second stage of electronic model construction is a classical picture that assigns physical meaning
 to $\mu_0$ and $C$ in Eq. (\ref{model1}).
We assume that heavily damaged GaAs behaves electronically like a featureless metal
 that is characterized by its Fermi level, $E_F$.
This assumption is based on the fact that Ga is a metal and As is a metalloid as elemental solids,
 and the forbidden energy gap only emerges from crystalline order.
The remaining crystalline GaAs is modeled as a featureless dielectric medium
 with a static dielectric constant of $\epsilon = 12.9$ \cite{GaAs_eps}.
Geometrically, we assume damage clusters to be spheres characterized by their radius, $R$.
The cluster model is based on the energy of a charged metal sphere embedded in a dielectric,
\begin{align}\label{model2}
 E(Q) &= - E_F Q + \tfrac{1}{2 \epsilon R} Q^2 , \\
 \Delta^Q_{Q+1} &= E_F - \tfrac{1}{\epsilon R} (Q + \tfrac{1}{2}) \notag ,
\end{align}
 in Hartree atomic units.
From this model, we hypothesize that $\mu_0 = E_F$ is a constant for damage clusters in the high-damage limit,
 and $C = \epsilon R$ is specific to each cluster and proportional to the cube root of its volume.

%P3.5 - Amorphous GaAs with a subgap, 3 parameters
The third stage of electronic model construction retains the simple geometric picture of the second model
 and refines the electronic behavior of the damage clusters.
While sufficiently high-energy damage to a GaAs crystal should produce metallic behavior,
 low-energy amorphous GaAs is observed to retain semiconducting behavior
 with a reduced band gap \cite{aGaAs}.
In this case, the common Fermi level for electrons and holes in Eq. (\ref{model2})
 is replaced by valence and conduction band edge energies of the semiconducting damage cluster,
 $E_{\alpha V}$ and $E_{\alpha C}$, with an assumed alignment with respect to the
 crystalline band edges of
\begin{equation}\label{alignments}
 E_V \le E_{\alpha V} \le E_F \le E_{\alpha C} \le E_C.
\end{equation}
The updated cluster model is
\begin{align} \label{model3}
  E(|Q|) &= - E_{\alpha V} |Q| +  \tfrac{1}{2 \epsilon R} Q^2, \\
  E(-|Q|) &= E_{\alpha C} |Q| +  \tfrac{1}{2 \epsilon R} Q^2, \notag \\
 \Delta^{|Q|}_{|Q|+1} &= E_{\alpha V} - \tfrac{1}{\epsilon R} (|Q| + \tfrac{1}{2}) \notag , \\
 \Delta^{-|Q|-1} _{-|Q|} &= E_{\alpha C} + \tfrac{1}{\epsilon R} (|Q| - \tfrac{1}{2}) \notag ,
\end{align}
 which continues to assume that all excess charge lies on the surface of the sphere as in a metal.
While $E_F$ is assumed to be universal for a highly damaged region of GaAs,
 $E_{\alpha V}$ and $E_{\alpha C}$ should vary between clusters and approach
 $E_{V}$ and $E_{C}$ as damage is annealed and the semiconducting crystalline state recovers.

%P3.6 - Simple application: distribution of identical point defects (Ga antisite)
The assumption of spherical geometry does not reduce the generality of the cluster model.
We demonstrate this by assuming the damage cluster to be a distribution of point charges
 that have a continuously tunable positive charge, $q_i$.
For a set of interpoint distances, $r_{ij}$, the minimum Coulomb energy of this cluster
 with a total positive charge of $Q$ is
\begin{equation}
 \min_{\substack{q_i \ge 0 \\ Q = \sum_i q_i }} \sum_{i \neq j} \frac{q_i q_j}{2\epsilon r_{ij}},
\end{equation}
 which is a quadratic programming problem.
We rescale the variables to $x_i \equiv q_i/Q$ and reduce the problem to
\begin{equation}
 \tfrac{1}{2 \epsilon} Q^2 \underbrace{\min_{\substack{x_i \ge 0 \\ 1 = \sum_i x_i }} \sum_{i \neq j} \frac{x_i x_j}{r_{ij}}}_{1/R},
\end{equation}
 with a solution that defines an effective spherical radius for an arbitrary spatial distribution of point charges.

\section{Results\label{result_section}}

%P4.1 - Overview of results
We validate and calibrate the damage cluster model with the combined results of MD and DFT simulations.
The limits on time and length scales of direct simulation necessitate finite-size scaling assumptions
 to provide a complete physical picture.

%F1 - DOS of amorphous versus crystalline GaAs (HSE06)
\begin{figure}
\includegraphics{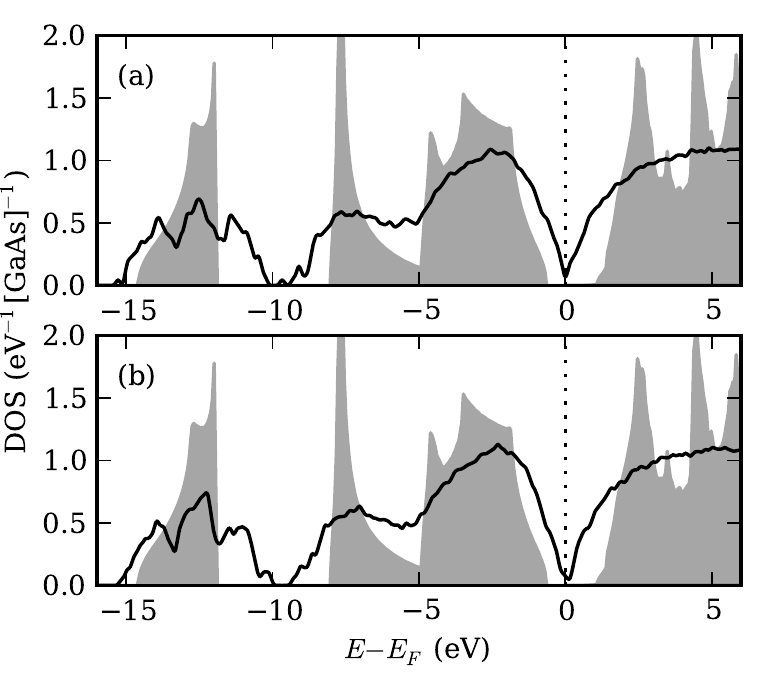}
\caption{\label{fig_DOS}Electronic density of states of crystalline GaAs (gray shaded region) compared to amorphous GaAs structures.
The structure in (a) is formed by annealing at 2500 K. The structure in (b) is the first structure with As-As and Ga-Ga bonded minimized
 by pairwise atomic rearrangements and further annealing at 800 K.
The Fermi level of the amorphous structures is aligned to the center of the crystalline GaAs band gap.}
\end{figure}

%P4.2 - Verification of metallic behavior in annealed damage
We assume that the two MD-generated bulk amorphous GaAs structures are representative of the interior of large
 damage clusters that cannot be simulated directly with DFT.
Contingent on this assumption, we verify that heavily damaged GaAs behaves electronically like a metal,
 as observed in previous studies \cite{GaAs_amorph_energy}.
The electronic DOS of the two structures are plotted in Fig. \ref{fig_DOS}
 against the electronic DOS of semiconducting crystalline GaAs.
There is little variation between the two cases, suggesting an approximately universal electronic DOS
 characteristic of heavily damaged GaAs.
Both samples are verified to be metallic,
 but with a strong pseudogap behavior that produces a DOS minimum at the Fermi level.

%F2 - Fit to capacitor model
\begin{figure}
\includegraphics{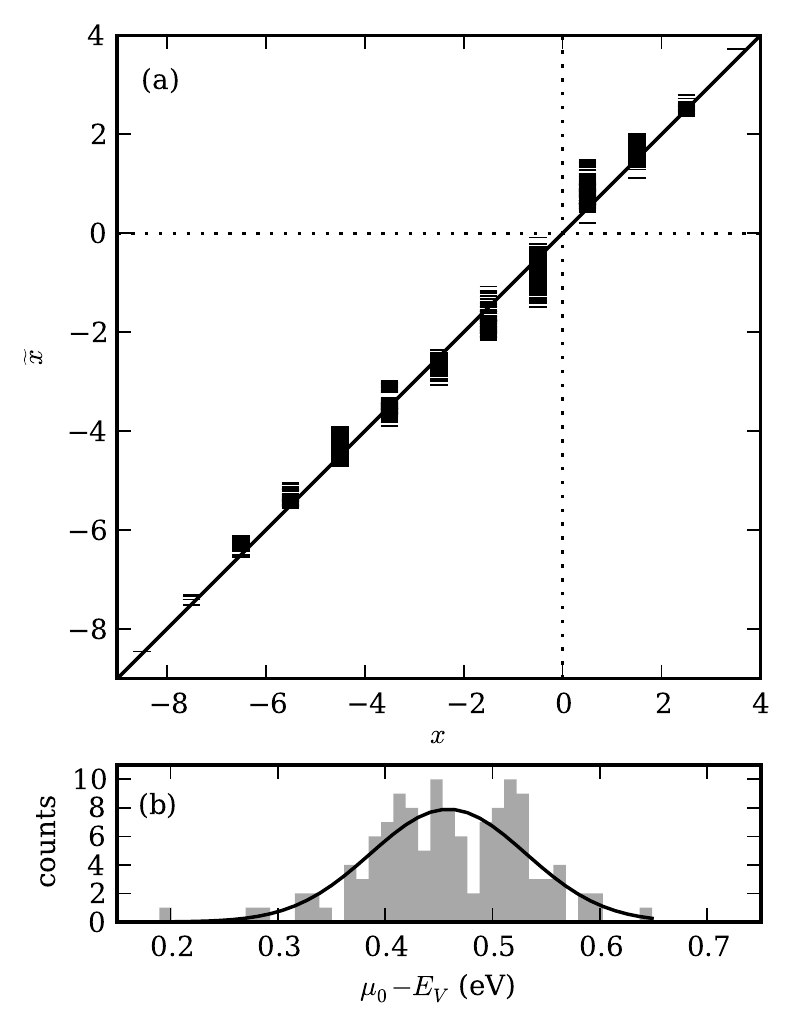}
\caption{\label{fig_cap} Artificial damage clusters fit to a capacitive model of their defect levels, $\Delta_{x+0.5}^{x-0.5}$.
(a) The correspondence between $x$ and the fit value, $\tilde{x}$, defined in Eq. (\ref{xfit}).
(b) The distribution of $\mu_0$ values from the optimized models and a Gaussian fit.}
\end{figure}

%F3 - Fit to metal sphere model
\begin{figure}
\includegraphics{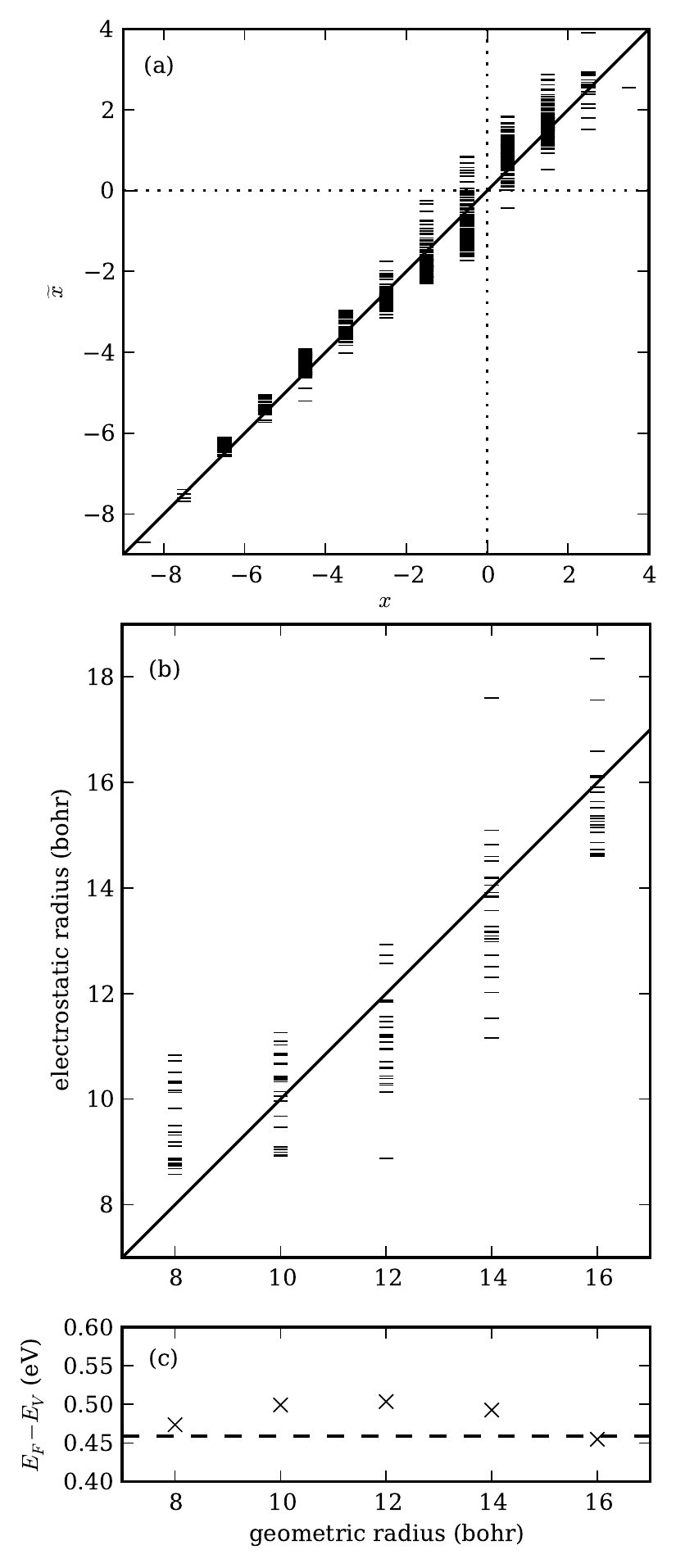}
\caption{\label{fig_fit} Artificial damage clusters fit to a metallic sphere model of their defect levels, $\Delta_{x+0.5}^{x-0.5}$.
The correspondence (a) between $x$ and the fit value, $\tilde{x}$, defined in Eq. (\ref{xfit}), and
(b) between geometric and electrostatic radii.
(c) The variation of $E_F$ with cluster size plotted against the globally optimized value (dashed line).}
\end{figure}

%P4.3 - Model fits to artificial clusters
We estimate the Fermi level of heavily damaged GaAs by fitting the models developed in Section \ref{model_section}
 to the vertical defect levels of the artificial damage clusters described in Section \ref{method_section}.
In Fig. \ref{fig_cap}a, each cluster is independently fit to the capacitive model in Eq. (\ref{model1}).
Each calculated defect level $\Delta_{x+0.5}^{x-0.5}$ is labeled by $x$, and $\mu_0$ and $C$ are determined by
 a least-squares minimization of $\tfrac{1}{C}(x - \tilde{x})$ with
\begin{equation}\label{xfit}
 \tilde{x} \equiv C ( \mu_0 - \Delta_{x+0.5}^{x-0.5}).
\end{equation}
The optimized values of $\mu_0$ are approximately normally distributed about $E_V + 0.46$ eV with a standard deviation of 0.07 eV,
 as shown in Fig. \ref{fig_cap}b.
This supports the assumption that the electronic properties of these clusters
 derives from a common metallic state with a well-defined Fermi level.

%P4.4 - Independent least-squares fit determination of E_F
In Fig. \ref{fig_fit}a, the clusters are fit to the metal sphere model in Eq. (\ref{model2}),
 by least-squares minimization of $\tfrac{1}{\epsilon R} (\tilde{x} - x)$ and a common value of $E_F$ for all clusters.
The optimized value of $E_F$ is 0.46 eV, in agreement with the mean of the $\mu_0$ distribution of the previous fit.
With fewer overall free parameters, the root-mean-square (RMS) error in $\tilde{x}-x$ grows from 0.31 to 0.43.
The clusters are constructed with a well-defined geometric radius
 which should be comparable to the radius, $R$, contained in the metal sphere model.
The geometric radius is consistently larger than $R$, which we attribute to a finite skin depth of the metal, $R_\mathrm{skin}$.
A least-squares fit of the electrostatically derived radius, $R + R_\mathrm{skin}$, to the geometric radius
 produces a skin depth of 2.1 bohr and an RMS error of 1.3 bohr, shown in Fig. \ref{fig_fit}b.
$R_\mathrm{skin}$ is in reasonable agreement with the GaAs skin depth of 1.6 bohr used to extrapolate the total energy of charge defects.
To assess finite-size errors of the $E_F$ estimate,
 the metal sphere model is refit to subsets containing clusters of fixed geometric radius.
No significant dependence of $E_F$ on radius is detected in Fig. \ref{fig_fit}c.

%F4 - Radial distribution of charge in the defect clusters
\begin{figure}
\includegraphics{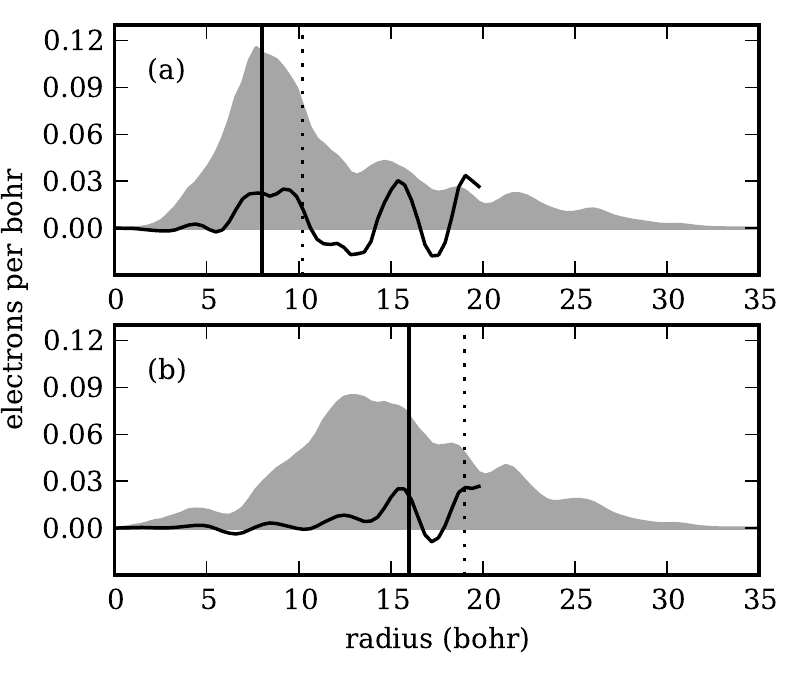}
\caption{\label{fig_rad} Radial distribution of excess charge in two damage clusters with one electron added to each.
We compare the charge density of the defect state (gray shaded region) to the total electronic charge rearrangement between the charged and neutral clusters (line) and the geometric radius (vertical solid line) to the electrostatic radius (vertical dotted line).
The electrostatic radius is defined here as the minimum radius containing the total excess charge of magnitude $\epsilon^{-1}$
 that remains after screening from bulk GaAs.}
\end{figure}

%P4.5 - Spatial distribution of charge in the damage cluster
We further validate the metal sphere model by visualizing the distribution of excess charge
 within damage clusters in Fig. \ref{fig_rad}.
One each of the smallest (8 bohr) and largest (16 bohr) radius artificial clusters is selected randomly
 and one electron is added to the neutral state.
In each case, the defect state that the excess electron occupies is delocalized over the cluster
 and has a significant tail extending outside the cluster.
Because of the large polarizability within the metallic cluster, charge fluctuations are heavily
 screened from the interior and a net screened charge of magnitude $\epsilon^{-1}$
 is confined to within a few bohr of the geometric radius of the clusters,
 in agreement with the skin depth fit to the metal sphere model.
Charge rearrangement beyond 20 bohr from the center of the clusters is omitted from Fig. \ref{fig_rad}
 because it contains finite-size artifacts.
The LMCC method removes the effect of these artifacts from the total energy,
 but they remain in the electronic charge density at the edge of the defect-centered Wigner-Seitz cell.

%P4.6 - Energy validation of artificial amorphous clusters
The artificial damage clusters and bulk amorphous structures have related metallic electronic properties,
 which is assumed to derive from sufficient damage to the crystalline structure.
We assess the degree of damage in these structures by the excess cohesive energy per atom of the damaged regions relative to crystalline GaAs.
For amorphous GaAs, typical numbers are 0.59 eV/atom for a highly disordered metallic structure
 and 0.36 eV/atom for an annealed semiconducting structure \cite{GaAs_amorph_energy}.
The bulk amorphous GaAs structures that we have generated have excess energies of 0.61 eV/atom and 0.58 eV/atom.
With only atoms inside the bounding spheres counted as damaged, the artificial clusters have an average excess energy of 0.74 eV/atom
 with an RMS deviation of 0.12 eV/atom between clusters.
By this criterion, the clusters are more damaged than the bulk amorphous structures,
 which is consistent with the absence of an MD-based annealing step in their creation.
Physical damage clusters should also have a high excess cohesive energy
 because of kinetic limitations to damage annealing.

%F5 - Defect levels of the 10 MD-generated clusters
\begin{figure*}
\includegraphics{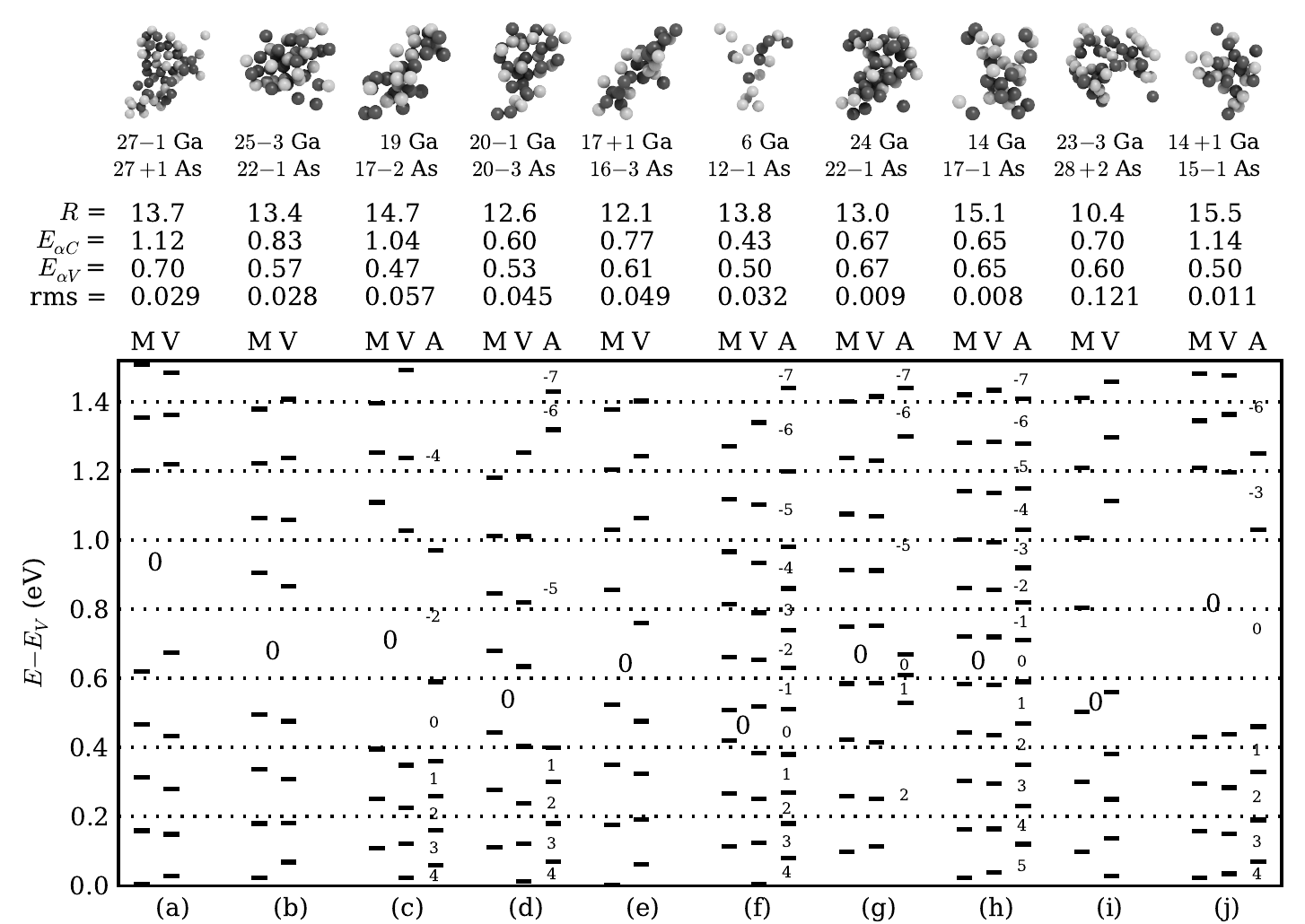}
\caption{\label{fig_level} 10 GaAs damage clusters extracted from MD simulations, with (h-j) further annealed after extraction ((h) is the annealed version of (g)).
Structures are visualized in \textsc{QuteMol} \cite{qutemol}, with Ga and As as black and white spheres.
Only atoms that deviate from their ideal crystalline positions by more than $5 \%$ of the lattice constant are shown.
Total number of defective atoms in the cluster are given, with $+/-$ denoting the deviation from the crystal stoichiometry.
The fit to the semiconducting sphere model in Eq. (\ref{model3}) as well as the RMS error in $\Delta_{Q+1}^{Q}$ is given for each cluster.
Transitions between stable charge states, $\Delta_{Q'}^{Q}$, are shown along with the $Q$ value of the energy intervals for the model (M), vertical (V), and adiabatic (A)
 calculations of cluster energies. $Q' = Q+1$ for all model and vertical charge states, and only the $Q=0$ charge state is labeled.}
\end{figure*}

%P4.7 - Initial discussion of the MD-generated clusters
There are not enough MD-generated damage clusters to produce useful statistics,
 but we examine them individually to identify common behaviors.
The details of these clusters are summarized in Fig. \ref{fig_level}.
The vertical defect levels of the clusters are fit to the semiconducting sphere model in Eq. (\ref{model3}).
There is only one significant outlier in this fitting process, Fig. \ref{fig_level}i.
This structure has a large energy window of stability for the $Q=-1$ charge state
 instead of the neutral state as predicted by the semiconducting sphere model.
We interpret this as a hole doped into the cluster, resulting from the overall excess of 5 As atoms.
By adjusting for this in the model, the fit improves to an RMS error in the energy levels of $0.025$ eV.
5 of the 10 clusters exhibits a band gap, $E_{\alpha C} - E_{\alpha V}$, greater than 0.2 eV,
 which validates the need for a generalization of the metallic sphere model.
The assumed energy alignment in Eq. (\ref{alignments}) is satisfied within the uncertainty of $E_F$.

%P4.8 - Adiabatic transitions & correlation of relaxations with energy of damage / degree of annealing
The lifetime of a charged state is likely to be longer than the timescale of structural relaxation.
Therefore, the adiabatic defect levels are more physically relevant than the vertical levels.
However, the relaxation of charge states exacerbates deficiencies in MD-based annealing of the damage clusters.
MD simulation does not account for electronic charge fluctuations,
 and only the energy surface of the neutral structure is sampled.
There can exist structural relaxations that significantly and irreversibly lower the energy
 with a large energy barrier in the neutral state, but with a small barrier or none at all for a charged state.
A more physical annealing process would enable charge fluctuations
 that significantly increase the transition rate into such lower energy structures.
4 of the 10 MD-generated clusters have unstable charge states that cause irreversible structural relaxations.
In two cases, Figs. \ref{fig_level}f and \ref{fig_level}h, relaxation effects produce a small perturbation in the defect levels.
In the remaining cases, relaxation effects are large and
 cause many charge states to be unstable at all values of the chemical potential.
While MD-based annealing does not prevent the charging instability of the cluster in Fig. \ref{fig_level}i
 or the large relaxation effects of the cluster in Fig. \ref{fig_level}j,
 it does eliminate the large relaxation effects of the cluster in Fig. \ref{fig_level}g
 by transforming it into the cluster in Fig. \ref{fig_level}h.
This evidence is not conclusive, but it demonstrates that MD-based
 annealing is capable of reducing relaxation effects but incapable of reliably relaxing unstable charge states.
Relaxation effects were not studied for the artificial damage clusters.
We conjecture that more thorough annealing would significantly reduce relaxation effects on average
 and that our study of vertical defect levels is representative of the adiabatic defect levels of clusters
 that have undergone such an annealing process.

\section{Discussion\label{discuss_section}}

%P5.1 - Extrapolation of results
The results of the previous section clarify that direct MD-plus-DFT simulation of damage clusters
 in GaAs are limited by both the size of clusters studied in DFT and the timescale of clusters annealed in MD
 combined with a lack of charge fluctuations in MD.
However, the electrical charging of even very small clusters are accurately fit to a classical model,
 and we expect the model to continue to work for larger, ``more classical'' damage clusters.
The model contains a geometric factor that can be estimated without quantum mechanical simulation
 and amorphous semiconductor band edges that asymptote to a universal Fermi level in the high-damage limit.
What remains unknown are the initial population and size-distribution of damage clusters
 as a function of incident ion energy, the long-time-dependence of cluster annealing,
 and the change in electronic properties as damage within the cluster is repaired by annealing.
Characterization of the initial distribution of damage clusters will be reported in a separate paper \cite{Foiles_prep}.
The remaining unknowns require further work to completely clarify.

%P5.2 - What do models based on point defects get right?
It is worthwhile to compare our partially constructed model of damage clusters with the existing radiation damage model of point defects.
A key difference is that the number of defect levels in a dilute distribution
 of point defects is proportional to the number of damaged atoms,
 while the number of levels in a damage cluster is dependent on the volume of the cluster and
 insensitive to the number of damaged atoms, as apparent in Fig. \ref{fig_level}.
Superficially, this is a significant difference between models.
However, the point defect model adjusts the energy levels of the point defects with the local electrostatic potential,
 which captures the same basic effect as the electrostatic charging energy
 of the models developed in Section \ref{model_section} for a dense distribution of charged points defects.
While damage clusters cannot be structurally decomposed into well-defined point defects in general,
 it may be possible to produce the same set of defect levels with a carefully chosen high-density distribution of point defects.
With such an approximation scheme, the electronic behavior of a multiple-point-defect model is able to mimic cluster behavior.

%P5.3 - What do models based on point defects get wrong?
Because a radiation damage model based on only point defects is capable of approximating the electronic properties of damage clusters,
 fitting such a model to experiments may prescribe an excess population of point defects to produce the same electronic behavior as the missing clusters.
Whether or not such an effective model is accurate depends on the difference in time-dependence between
 a dense distribution of point defects and a damage cluster.
The rate of binary chemical reactions between point defects will increase with density,
 which might reduce the lifetime of a dense distribution of point defects relative to a damage cluster.
The long-time evolution of a damage cluster is not yet available for comparison.
Also, the density of a distribution of point defects can be dynamically reduced either by
 diffusion or by drift resulting from Coulomb repulsion between charged point defects.
These are dynamical degrees of freedom that a damage cluster is not likely to have,
 assuming some degree of cohesion within the cluster and negligible mobility of the cluster as a whole.

%P5.4 - What more information is necessary to build a complete damage model including clusters ?
We propose a simple rationalization of the time evolution of a damage cluster,
 which directly relates to the parameters of the semiconducting sphere model in Eq. (\ref{model3}).
First, we assume that the amorphous GaAs band edges are directly correlated
 to the average excess cohesive energy per atom over the crystalline state.
Next, we assume that the effect of annealing can be decomposed into a reduction of the excess cohesive energy
 plus a surface recrystallization that is nucleated by the crystalline region outside the cluster.
These effects are a rationale for the distinct time-dependence of $E_{\alpha V}$ and $E_{\alpha C}$ and of $R$.
The rate of excess cohesive energy reduction and the speed of the crystallization front
 can be studied separately in idealized simulations of amorphous GaAs or of an interface between crystalline and amorphous GaAs.
In addition to the evolution of the damage cluster, there also may be dynamic interactions
 between clusters and point defects.
For example, damage clusters might act as a source or sink for point defects.
The exploration of these ideas is left to future work.

\section{Conclusions}

%P6.1 - What we did and what remains to be done
A complete radiation damage model in GaAs must account for the initial conditions,
 electronic properties, and time-dependence of all relevant material defects following a radiation event.
It has been assumed previously that all relevant defects are simple, enumerable point defects,
 but we have established the commensurate relevance of larger damage clusters.
The defect energy levels of all damage clusters cannot be exhaustively computed,
 but they are well described by a classical total energy model fit to a statistical sampling of damage clusters.
This result highlights the importance of sampling a statistical distribution when
 studying physical phenomena involving disorder at the atomistic scale.
A complementary report on the initial distribution of damage clusters observed in MD simulations of collision cascades is in preparation \cite{Foiles_prep}.
We have outlined a possible model of the long-time evolution of damage clusters,
 but validation and parameterization of the model remain as future work.
This remaining problem is difficult because of the long timescales involved and the possibility
 that electronic charge fluctuations are an important relaxation mechanism
 not present in tools such as MD that are capable of directly accessing the necessary time scales.

\begin{acknowledgments}
Sandia National Laboratories is a multi-program laboratory managed and  
operated by Sandia Corporation, a wholly owned subsidiary of Lockheed  
Martin Corporation, for the U.S. Department of Energy's National  
Nuclear Security Administration under contract DE-AC04-94AL85000.
\end{acknowledgments}

\end{document}